\documentstyle[12pt]{article}
\begin{document}
\title{Space-time encoding} \author{Arindam Mitra
\\Lakurdhi, Tikarhat Road, Burdwan, 713102,\\
 West Bengal,  India.}

\maketitle
\begin{abstract}It is widely believed that one signal can carry maximum
one bit of information, and minimum KTlog2 energy is required
to transmit or to erase a bit. Here it is shown four bits of information
can be transmitted by one signal  in four dimensional space-time. It implies space-time
cannot be ignored to determine the minimal energy cost of information
processing.
\end {abstract}
\newpage
\noindent

Before the turn of the 19th century, Bell's invention of telephone and 
J. C. Bose's
first demonstration of radiowave transmission  heralded  a new era of information
transfer. In the following years, although these new techniques of almost instantaneous
information transmission were rapidly developed and widely implemented, but it was not known how to quantify information. In 1940s, Shannon [1]  proposed entropy as  a measure of information, which has been
accepted by all. So, information is not an abstract thing, it can be expressed in terms of
entropy which is basically a thermodynamical quantity. Moreover, information is
carried by electromagnetic signal. Due to these reasons many authors conjectured
that the laws of physics would determine the fundamental issues of information
processing. In favor of this belief, one can say information cannot be
transmitted faster than the velocity of light. But for other issues, the connection
between physics and information theory is yet to be understood.
As for example, we don't know how to incorporate space-time with information
processing. In the existing encoding space-time has no explicit role.
As information is processed within the space-time, one can look for a direct connection
between the two. In this paper we shall try to establish such connection.\\

In  information theory,  the basic unit of information is called
bit, which is represented by an electromagnetic signal. Now, the question is
: How many bits can be transmitted by a signal ? The general wisdom is that not more than one
bit can be transmitted by a signal, however there is no fundamental reason why it would be so.\\

Let us judge a signal according to special theory of relativity.
According to special theory of relativity,
an event needs four dimensions for its complete specification - of which 
three are space dimensions
and one is time dimension. As information carrying signal represents an event, so we can tell that
a signal requires four bits of information for its complete
specification. So, it is natural to think that four bits of 
information can be transmitted by a signal. Next we shall describe how it is possible.\\

Suppose a three dimensional picture is transmitted, which carries the bit
values. As a naive method of transmission, we can think that this three
dimensional signal is projected on a screen of the receiver, who by
measuring /observing the picture can recover the bit values. Suppose
$\delta x_{i}$, $\delta y_{i}$ and  $\delta z_{i}$ are length, breadth
and height of the picture. Now $\delta x_{i}$  can take any of the two values
$\delta x_{o}$ and  $\delta x_{1}$ which represent bit 0 and 1 respectively.
Similarly $\delta y_{i}$ can be $\delta y_{0}$ or  $\delta y_{1}$ and
$\delta z_{i}$ can be $\delta z_{0}$ or $\delta z_{1}$. So each figure contains three bits simultaneously. The probable
sets of three bits are: (0,0,0), (0,0,1), ( 0,1,0), (0,1,1), (1,0,0), (1,0,1)
(1,1,0) and (1,1,1).  These triple bits are generated due to three dimensional space.
We can call these bits as space-bits. Let us see how time dimension can be 
incorporated to generate another bit
from the same picture. Suppose sender transmits the picture over the time period
$\delta t_{i}$, where $\delta t_{i}$ can take any two different values
 $\delta t_{0}$ and  $\delta t_{1}$. Let us call them time bits. 
The two values of time duration  can
represent bit 0 and 1 respectively. So each figure contains 4 bits of which three are
space bits and one is time bit.  Therefore information can be encoded into the 
16 possible sets of 4 bits. The probable sets are: ( 0,0,0,0), (0,0,1,0), (0,1,0,0),
(0,1,1,0), (1,0,0,0)(1,0,1,0),(1,1,0,0), (1,1,1,0),(0,0,0,1), (0,0,1,1),
(0,1,0,1),(0,1,1,1), (1,0,0,1), (1,0,1,1), (1,1,0,1), (1,1,0,1). 
These bits can be
called as space-time bits.\\

Is it possible to transmit four spec-time  bits by the existing information processing technique ?
In the existing technique, an electromagnetic/voltage pulse carries
the bit value. In three dimensional configuration, voltage pulse $v$ can be expressed as
$ v = (v_{x} i + v_{y} j + v_{z} k)^{1/2} $ where i, j and k are three unit
vectors along x, y and z axis. To recover the bit values receiver measures
the magnitude $ (v_{x}^{2} + v_{y}^{2} + v_{z}^{2})^{1/2}$ of the voltage pulse.
Note that he measures the resultant of three components of voltage pulse as they
have no independent existence. Therefore receiver can extract only
one bit of information by measuring the voltage of the voltage pulse.  
It means
sender can dump only one bit of information into
three dimensional voltage pulse. But time bit can be sent along with the
voltage pulse.  Let  voltage $v_{0}$ and  $ v_{1}$  represent
two bit values. Suppose the time width of the pulse can be either $\delta t_{0}$
and $\delta t_{1}$ respectively 0 and 1. Therefore, receiver can recover two pulses
by measuring the voltage and time duration of the pulse
provided $v_{0}\neq 0$ and $v _{1} \neq 1$. The probable sets of two bits are :(0,0),
(0,1), (1,0) and (1,1). In this way, two bits can be transmitted by 
one voltage pulse. \\

 The presented space-time encoding is perhaps a classical encoding since
 space-time has no quantum analogue.  Over the last few years the  quantum information has been
 a major area of research [3, 4]. In quantum information, quantum state is the carrier of
 information where a single quantum state or a sequence of quantum states [3,4]
 can represent a bit value. So the same question can be put:
 Is it possible to realize our space-time encoding in  quantum information ?
 Suppose eigen values of spin state are used to represent bit 0 and 1.
 The state  have three eigen values in the x, y and z axis, 
 but they cannot be
 simultaneously measured. Therefore it is not possible to 
encode more than one bit by a
spin state.   In this sense,
 three dimensional space has no advantage in quantum encoding.
 So voltage pulse and quantum
 state are equivalent in this sense. Still they have a difference.
 We have seen one voltage pulse can transmit  2 bits. But one
 quantum state cannot transmit more than one bit.  The reason is simple.
 In quantum encoding we can use either two orthogonal states or two nonorthogonal
 states to represent two bit values. But  nonorthogonal states
 cannot generate bit for each state
 (the probability of recovery of a bit  value is less than one), only orthogonal states can do so, . So we are interested to know
 whether two bits per quantum state can be generated by two orthogonal states.  For clarity, suppose
 $ \vert \updownarrow\rangle$
  and $\vert \leftrightarrow\rangle$ are two orthogonal polarization
  states, which represent 0 and 1. Receiver uses $0^{\circ}$ or $90 _{\circ}$
  analyzer to recover bit values.
  Therefore, receiver could  recover 100\% transmitted bits if time of transmission is
  a prior known. Note that 50\% bit will be recovered completely from null results. 
     These 50\% null results cannot provide more than 50\% bit values. It means
     2 bits per quantum state cannot be generated. Four dimensional space-time
has no advantage in quantum encoding.\\

    Throughout the development of classical information theory, many authors tried to establish
    a relation between thermodynamics and information theory.  In particula, 
    they
    tried to know the minimal energy cost of information processing. von Neumann [3] and Brillouin [4]
    argued that KT log 2 energy is required to process a bit. Gabor argued [5] transmission
    of one bit of information by electromagnetic
    waves of radio frequency requires a minimum energy KT. On the other hand
    Shannon formula of channel capacity [1] states that average energy requirement of one bit transmission
    is KTlog 2.  Gabor's and Shannon's results are not identical
    because Shannon formula is not applicable to the transmission of one-bit of information.
    At present it is widely believed that KTlog2 is  required to transmit a bit.
    But from the earlier work it is not clear whether
    this energy is dissipative in nature or not.  On the question of dissipation,
     Landauer
    argued [6] that KTlog2 would dissipate whenever a bit will be erased, otherwise not.
    Erasure of information is frequently done in computation.
    Following Landauer's argument
    Bennett argued [7] that {\em computation} can be done without spending any energy when
    computation does not require any erasure of information. This new model of
computation is  known as  reversible computation. From these works
    one may think    dissipation-less information processing is possible only
    when it is processed reversibly. In a separate work Landauer proposed an
    alternative encoding [8] and argued that
    one bit of information could be transmitted without any energy dissipation.
   It means reversibility is not a necessary condition for dissipation-less information processing.
 But, according to Landauer's statement [9] his work created little attention.  In true
   sense, Landauer's scheme is not a bit transmission scheme, rather it is a
   bit transportation scheme. Therefore one can think KTlog2 result is valid for
   erasure of bit and transmission of bit.\\

   Many authors [10] cast doubt either on dissipation-less computation or
   on the issue of energy requirement of erasing a bit
   because of the lack of  mathematical proof.
   Apart from the lack of  proof,  KTlog 2 result is based on the assumption
   that a signal contains only  one bit of information. We have seen
   a signal can contain four bits of information. The transmission of this
   signal or erasing the signal 
   requires KTlog2 energy (assuming the result  is valid). The energy 
  cost of transmission
  of four bits or erasing the four bits is KTlog 2.
    It means
    encoding cannot be overlooked to determine the energy requirement of bit
    processing. It also implies  space-time will have a  role
    on the fundamental limit's of energy cost of classical information processing.
    But for quantum information, space-time will not have  such
    similar role because it is not possible to send more than
    one bit by a quantum state. \\

    In conclusion, information bearing capacity of space-time has been demonstrated.
    In this encoding space and time have been used on equal footing. 
On the other hand this  encoding  qualitatively supports the notion of four dimensional space-time.
    We hope connection between space-time and information theory might be an 
interesting topics
    of research.\\


\begin{thebibliography}{99}  

\bibitem {1} C. Shannon, Bell System Technical Jour, (1948).

\bibitem{2}A. Mitra. quant-ph/9812087;  physics/0007079 ; See reference.

\bibitem{3}von Neumann {\em Theory of Self Reproducing Automata} ( Univ. Illiuonis Press, Urbana , 1960).

\bibitem{4}L. Brillouin {\em In science  and information  Theory}
Chap. 13, p. 162,

\bibitem{5}D. Gabor, Phil. Mag {\bf 41}, 1161 (1950).

\bibitem{6}R. Landauer, IBM. J. Res. Dev  {\bf 5}, 183 (1961).

\bibitem{7}C. H. Bennett, IBM J. Res. Dev, {\bf 17}, 525 (1973)

\bibitem{8}R. Landauer , Appl. Phy. Lett, 51, 2056 (1987)

\bibitem {9} R. Landauer, Phy. Lett A. {\bf 217}, 188 (1996)

\bibitem{10} W. Porod, O. R. Grondin, K. D. Ferry, G. Porod, Phy. Rev Lett,
{\bf 52}, 232, (1984); D. Wolpart, Phy. Today {\bf 45 }(3), 98, 1992;
E. Goto, W. Hioe, M. Hosoya, Physica, C, 385 (1991);
K. Shizume phy. Rev. E 52, 3495 (1995).

           

\end{thebibliography}
\end{document}